**Fundus2Globe: Generative AI-Driven 3D Digital Twins for Personalized Myopia Management**

Danli Shi, MD, PhD[1,2#], Bowen Liu, MS[1#], Zhen Tian, MS[1], Yue Wu, MD[1], Jiancheng Yang, PhD[3], Ruoyu Chen, MD[1], Bo Yang, PhD[4], Ou Xiao, MD, PhD[5], Mingguang He, MD, PhD[1,2,6*]

**Affiliation**

1. School of Optometry, The Hong Kong Polytechnic University, Kowloon, Hong Kong

2. Research Centre for SHARP Vision (RCSV), The Hong Kong Polytechnic University, Kowloon, Hong Kong

3. Swiss Federal Institute of Technology Lausanne (EPFL), Lausanne, Switzerland.

4. Department of Computing, The Hong Kong Polytechnic University, Kowloon, Hong Kong

5. State Key Laboratory of Ophthalmology, Zhongshan Ophthalmic Center, Sun Yat-sen University, Guangdong Provincial Key Laboratory of Ophthalmology and Visual Science, Guangdong Provincial Clinical Research Center for Ocular Diseases, Guangzhou 510060, China.

6. Centre for Eye and Vision Research (CEVR), 17W Hong Kong Science Park, Kowloon, Hong Kong

#Contributed equally

*Correspondence

Prof. Mingguang He, Chair Professor of Experimental Ophthalmology, School of Optometry, The Hong Kong Polytechnic University, Kowloon, Hong Kong. Email:

mingguang.he@polyu.edu.hk

mingguang.he@polyu.edu.hk

**Abstract**

Myopia, projected to affect 50% population globally by 2050, is a leading cause of vision loss. Eyes with pathological myopia exhibit distinctive shape distributions, which are closely linked to the progression of vision-threatening complications. Recent understanding of eye-shape-based biomarkers requires magnetic resonance imaging (MRI), however, it is costly and unrealistic in routine ophthalmology clinics. We present Fundus2Globe, the first AI framework that synthesizes patient-specific 3D eye globes from ubiquitous 2D color fundus photographs (CFPs) and routine metadata (axial length, spherical equivalent), bypassing MRI dependency. By integrating a 3D morphable eye model (encoding biomechanical shape priors) with a latent diffusion model, our approach achieves submillimeter accuracy in reconstructing posterior ocular anatomy efficiently. Fundus2Globe uniquely quantifies how vision-threatening lesions (e.g., staphylomas) in CFPs correlate with MRI-validated 3D shape abnormalities, enabling clinicians to simulate posterior segment changes in response to refractive shifts. External validation demonstrates its robust generation performance, ensuring fairness across underrepresented groups. By transforming 2D fundus imaging into 3D digital replicas of ocular structures, Fundus2Globe is a gateway for precision ophthalmology, laying the foundation for AI-driven, personalized myopia management.

**Introduction**

Myopia, defined as a spherical equivalent refraction of ≤ −0.50 diopter, is a leading cause of blindness due to refractive errors and is a growing public health concern. Currently affecting nearly 30% of the global population, this figure is expected to rise to 50% by 2050, with 10% experiencing high myopia, presenting a significant global socioeconomic challenge.[1] Pathological myopia is associated with abnormal changes in eye shape, which can lead to complications such as posterior staphyloma, macular degeneration, and retinal detachment. Given its role in reflecting structural alterations, eye shape has been increasingly recognized as a potential biomarker for myopia.[2]

High-resolution 3D magnetic resonance imaging (MRI) and advanced image analysis techniques have enhanced our understanding of ocular shapes, particularly in highly myopic eyes. Prior studies have used MRI and optical coherence tomography (OCT) to reconstruct 3D eye shapes and quantify morphological changes, uncovering strong associations between eye shape and vision-threatening complications in pathological myopia.[3-10] However, OCT offers a limited view of the macular region and cannot model the entire eye shape, while MRI's high cost, limited accessibility, and time-intensive procedures make it impractical for routine clinical use.

Recent advancements in generative artificial intelligence (GenAI) have demonstrated remarkable cross-modal modeling capabilities in healthcare, enabling the learning of shared representations and mapping between different modalities. Cross-modal GenAI enables the transformation of data between different imaging modalities, improving image quality, reducing the need for invasive procedures, enhancing disease diagnosis, refining segmentation, generating sensitive biomarkers, augmenting datasets for rare diseases, and enabling the creation of digital twins for organs.[11-18] In the context of pathological myopia, generating 3D eye shapes as digital twins from the widely accessible color fundus photographs (CFP) and metadata—holds promise for enhancing our understanding of myopia pathophysiology and facilitating personalized modeling of disease progression prediction and treatment interventions.

To date, no prior work has utilized CFP and metadata to generate a 3D eye globe (EG). This study introduces Fundus2Globe, a diffusion model designed to generate the 3D shape of highly myopic eyes reconstructed from 3D MRI data. It uses metadata, such as axial length (AL) and spherical equivalent (SE), along with latent representations of CFP as input. Both internal and external validation demonstrates that the generated 3D EG maintains realistic shape distributions and accurately preserves vision-threatening fundus lesions of posterior staphyloma. Furthermore, our experiments show that the model can generate plausible 3D EG counterfactuals from a single CFP and arbitrary conditioning with clinical parameters, potentially serving as a digital twin for improving personalized myopia management.

## Results

### Dataset

A total of 197 eyes with paired CFP-MRI from 99 patients in the Zhongshan Ophthalmic Center Brien Holden Vision Institute (ZOC-BHVI) Guangzhou High Myopia Cohort Study were used to develop Fundus2Globe. The patients' ages ranged from 12 to 67 years, with a median age of 26.17 (interquartile range: 21.71, 40.18). Of the participants, 50.5% were male. Dataset characteristics are summarized in Table 1.

We performed external validation using the PALM dataset[19], which comprises 1200 fundus images with binary labels for pathological myopia.

**Table 1. Characteristics of participants for Fundus2Globe development.**

| Participants | Total (n = 99) | | | |
|---|---|---|---|---|
| **Age, Median (Q1,Q3)** | 26.17 (21.71, 40.18) | | | |
| **Sex, n (%)** | | | | |
| Male | 50 (50.5) | | | |
| **Eyes** | Total (n = 197) | OD (n = 98) | OS (n = 99) | p |
| **SE, Median (Q1,Q3)** | -10 (-12.25, -8) | -10 (-12.69, -8.06) | -10 (-11.75, -7.62) | 0.52 |
| **Staphyloma, n (%)** | | | | 0.787 |
| no staphyloma | 166 (84.3) | 82 (83.7) | 84 (84.8) | |
| macular-affected | 8 (4.1) | 5 (5.1) | 3 (3) | |
| non-macular-affected | 23 (11.7) | 11 (11.2) | 12 (12.1) | |
| **VA, Median (Q1,Q3)** | 1 (0.7, 1) | 1 (0.7, 1) | 0.9 (0.7, 1) | 0.725 |
| **AL, Median (Q1,Q3)** | 27.87 (26.84, 29) | 28.05 (26.85, 29.04) | 27.64 (26.82, 28.98) | 0.707 |
| **Vertical asphericity, Median (Q1,Q3)** | -0.19 (-0.24, -0.15) | -0.19 (-0.24, -0.15) | -0.19 (-0.24, -0.14) | 0.57 |
| **Horizontal asphericity, Median (Q1,Q3)** | -0.13 (-0.19, -0.07) | -0.13 (-0.2, -0.07) | -0.12 (-0.18, -0.07) | 0.648 |
| **Volume/mm^3, Median (Q1,Q3)** | 9736.47 (9128.09, 10861.94) | 9692.45 (9138.4, 10877.22) | 9798.23 (9137.32, 10824.96) | 0.818 |

SE=spherical equivalent, VA=visual acuity, AL=axial length, OD=right eye, OD=left eye

## Robust and Efficient 3D Eye Globe (EG) Reconstruction

The overview of proposed framework is illustrated in Fig. 1. To achieve robust reconstruction of 3D EG using high-resolution MRI, we face the challenge of ensuring reliable segmentation despite strong signal intensity distinctions in globe areas. To ensure the reliability of segmentation results while minimizing manual effort, we propose a 3D reconstruction strategy, depicted in Fig. 1a, utilizing the SAM (see Method for details). SAM is a promptable segmentation system capable of zero-shot generalization to new objects and images, eliminating the need for additional training.[20,21] Reconstructing 3D EG from MRI involves three steps: segmentation, reconstruction, and meshing. For segmentation, SAM automatically detects all objects in the 2D slices and aligns labels with the same semantics. A correction process then produces a voxel representation of the 3D EG. We use the marching cubes algorithm to extract the surface from the volumetric data. In meshing, filters of Poisson surface reconstruction and surface subdivision were applied to achieve the 3D shape reconstruction into a relatively smooth mesh format.

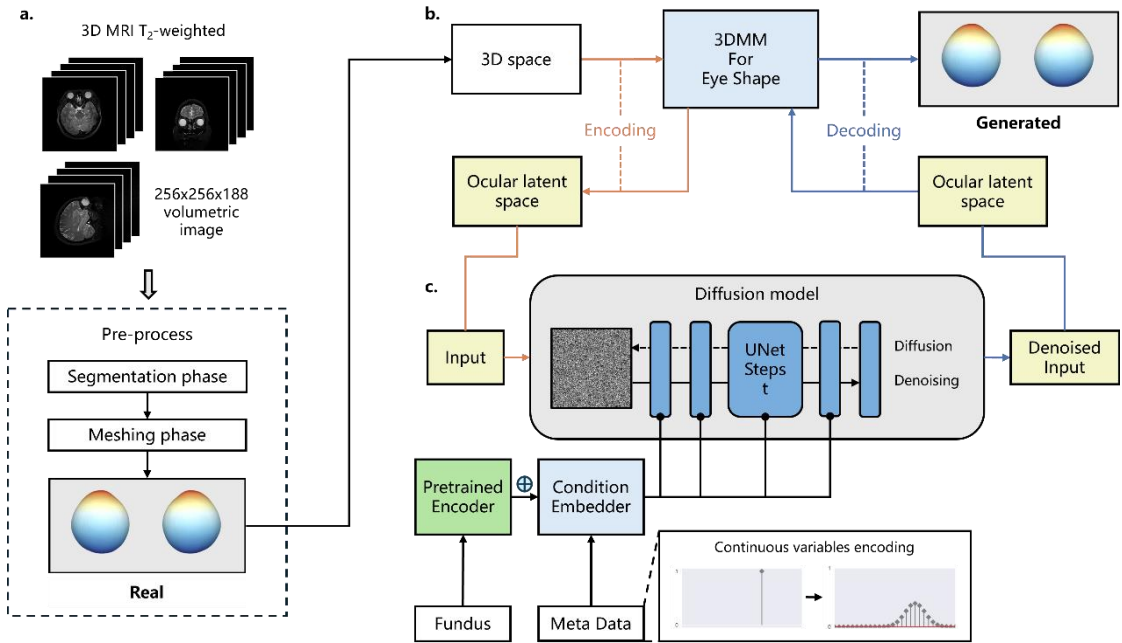

**Fig. 1: Fundus2Globe architecture for generating the 3D Eye Globe from color fundus photograph (CFP) and metadata.**

**a.** MRI pre-processing for 3D ocular shape reconstruction. This step involves segmenting 2D eye contour from T2-weighted MRI scans, which are optimized to highlight the fluid-filled chambers of the eye. The segmented slices are then used to reconstruct the ocular shape into a triangle mesh format, with colors applied for enhanced visualization. **b.** 3D Morphable Model (3DMM) for ocular shape representation. To incorporate a robust shape prior for efficient modeling, we developed a 3DMM based on our myopic dataset rather than relying on data-driven direct prediction. This specially designed morphable model enables the encoding and decoding of 3D eye globes (EG) by transforming them between 3D space and an ocular latent space. **c.** Fundus2Globe architecture for generating 3D EGs. It comprises a 3DMM as the eye shape encoder-decoder, a pre-trained retinal foundational model as the CFP encoder, and a Denoising Diffusion Probabilistic Model (DDPM) for cross-modal generation in latent space. During training, the CFP is encoded into a latent embedding, which is added with metadata embeddings—continuous variables encoded as descriptive vectors—to form the conditional embedding. To generate a new 3D EG, the process begins with complete random noise and iteratively generates a synthetic ocular latent guided by the conditional embedding. The generated ocular latent is then decoded into a 3D EG using the 3DMM.

One clinician reviewed the 2D segmentation results, identifying two instances of over-segmentation and one of under-segmentation across all 2D slices from the coronal, sagittal, and axial planes. After correcting these, the reconstructed 3D EG showed no shape anomalies due to reconstruction strategy failures, as confirmed by two clinicians. We showcased the segmentation results for MRI slices, 3D voxel masks, and the reconstructed 3D EGs in Supplementary Fig. 1.

The reconstructed 3D EG is stored in a triangle mesh format, which is compatible with most algorithms.[22-24] However, since the ocular shape is neither extremely complex nor completely irregular, using raw meshes for the diffusion model would significantly increase the data required for training. To address this, we developed a 3D morphable model based on our population data, as shown in Fig. 1b, to accurately represent the ocular shape of myopic eyes. This model facilitates the encoding and decoding of 3D EGs by transforming them between 3D space and a vector space, which we refer to as the ocular latent space in the following sections.

**Fundus2Globe architecture and experimental setting**

The Fundus2Globe architecture utilizes a latent diffusion model conditioned on multimodal inputs to generate anatomically plausible 3D eye globes (EGs). The model integrates metadata including, continuous attributes (axial length [AL], spherical equivalent [SE]) as well as discrete variables (eye laterality), and CFPs. High-dimensional CFPs, which exhibit structural homogeneity but localized feature diversity, are encoded into a compact latent representation using EyeFound,[25] a multimodal ophthalmic foundation model pre-trained to capture clinically relevant features while enabling robust generalization for multimodal tasks. This encoding step reduces computational complexity and adds fundus-derived features with metadata inputs.

To address challenges posed by the non-uniform spacing and incomplete value ranges of continuous attributes (AL, SE, VA), we adopted label distribution learning (LDL)[26].

LDL discretizes continuous values by converting them into a Gaussian distribution spanning multiple adjacent labels, thereby preserving similarity between neighboring values while mitigating data sparsity. Ablation studies compared LDL against two alternative approaches: raw continuous inputs (no discretization) and clustering-based discretization (aggressive grouping). LDL outperformed both extremes, balancing granularity and generalizability by incorporating inter-label relationships during training.

Model validation was conducted on internal and external datasets, with performance assessed globally and across subgroups stratified by posterior staphyloma status (none, macular-affecting, non-macular-affecting), age (<40 vs. ≥40 years), gender, and laterality. Topological fidelity between generated and real EGs was quantified using chamfer distance as the primary metric, supplemented by Hausdorff distance (and its one-sided variant[27,28]) and point-to-surface distance to measure spatial deviations[29]. . For all metrics, lower values reflect higher similarity. Additionally, a clinically relevant metric, the aspheric descriptor Q value[9]—defined as $(R_x / R_z)^2 - 1$, where $R_x$ and $R_z$ represent posterior eyeball width and length, respectively (Supplementary Fig. 2)—was compared between real and generated 3D EGs across staphyloma subgroups. This analysis evaluated the alignment of geometric distributions between synthetic and real EG under varying pathological conditions.

**Fundus2Globe accurately generates 3D EGs**

The main goal of Fundus2Globe is to accurately preserve the distribution of ocular shape and maintain abnormal shape characteristics of the 3D EGs based on the more easily accessible information provided. We first examined the primary metric, chamfer distance, under different experimental settings. Different experimental setups mainly reflect in the information used to guide the generation of the 3D EGs and its processing methods. To establish a more intuitive baseline, we applied a scaling transformation to the ground truth based on physical distance and calculated the chamfer distance between the ground truth and its transformed version in the 3D point cloud space. The selected scaling parameters vary from 0.5mm to 0.8mm (Fig. 2a).

Fundus2Globe demonstrated superior performance when guided by both CFPs and metadata compared to CFPs alone, particularly in capturing pathological shape variations (Fig. 2b). We believe this improvement is attributed to the complementary role of metadata in resolving ambiguities inherent to fundus images, such as subtle shape distortions caused by staphyloma. In terms of handling continuous sensitive attributes in metadata, LDL-based discretization reduced chamfer distance by 18% compared to raw continuous inputs and 12% versus clustering, highlighting its efficacy in handling sparse, non-uniform attributes. the LDL-based discretization method has a more significant advantage compared to the non-discretization method and the clustering-based discretization method (Fig. 2b). This is because LDL-based discretization considers not only the metadata values corresponding to the current 3D EG during training but also the learning between all adjacent values. In all the experiments, the processing of converting the latent embedding corresponding to the fundus and the left-right eye labels into category embeddings remained consistent (see Methods for details). We also performed test-time augmentation (TTA) to improve performance, i.e., repeatedly sampling different sub-samples under the same condition and averaging them as the final output. TTA provides additional performance improvement, though its impact was marginal relative to LDL's contribution (Fig. 2b). Notably, the model exhibited no performance disparities across age, gender, laterality, or staphyloma subgroups, underscoring its fairness in diverse clinical populations (Fig. 2c). Additional metrics, including the Hausdorff distance and its one-sided version, are presented in Supplementary Fig. 3.

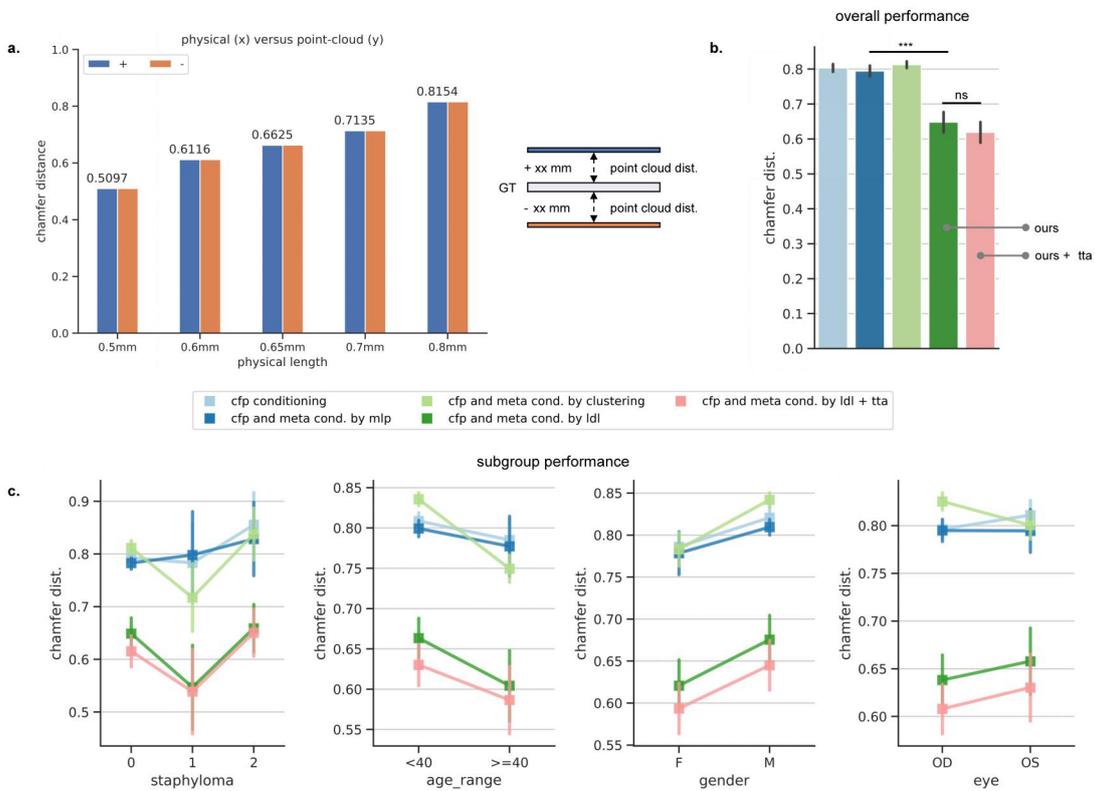

**Fig. 2: Generation quality of Fundus2Globe based on the Chamfer distance.**

a, Bar plot comparing the chamfer distance of the ground truth and its scaled version on different scaling parameters. The scaling transformation is based on physical length, with selected scaling parameters ranging from 0.5 mm to 0.8 mm.

b, Bar plot showing the chamfer distance of Fundus2Globe in comparison to competing methods. Data are presented as mean ± standard deviation. across n = 5 independent experiments. The following methods were compared: 'CFP conditioning' (conditioned solely on fundus images); 'CFP and meta conditioning by MLP' (conditioned on both fundus images and meta-data, where the variables in the meta-data are encoded by an MLP model); 'CFP and meta conditioning by clustering' (conditioned on both fundus images and meta-data, where continuous variables in the meta-data are grouped into clusters); 'CFP and meta conditioning by LDL' (utilizing label distribution learning-based discretization for encoding continuous meta-data variables); and 'CFP and meta conditioning by LDL + TTA' (adding test-time augmentation). LDL + TTA demonstrated superior performance across all other methods. The listed P-value indicates the statistical significance of Fundus2Globe outperforming the best comparison method, as determined by t-test.

c, Comparison of the chamfer distance of Fundus2Globe and competing methods across different subgroups (different classes of staphyloma, age, gender, and laterality). Data are mean ± standard deviation across 5 independent experiments.

Spatial accuracy was validated through point-to-surface distance metric, which measures the distance from each point on the generated EG to the surface of its realistic counterpart. Under the optimal configuration (LDL + TTA), the model achieved a root mean squared error (RMSE) of 0.2127 and a mean absolute error (MAE) of 0.2048 on the internal test set (Supplementary Fig. 4). Beyond distance-based evaluations, the generated EGs also preserved the Q-value distribution of real EGs across staphyloma subgroups, with no significant differences in asphericity ($p > 0.05$) (Fig. 3a). Staphyloma classification was defined as follows: 0 indicates no staphyloma, 1 indicates staphyloma not involving the fovea, and 2 indicates staphyloma involving the fovea. For qualitative analysis, we presented real and generated samples from each staphyloma subgroup (Fig. 3b, c, d) to visually demonstrate the model's performance.

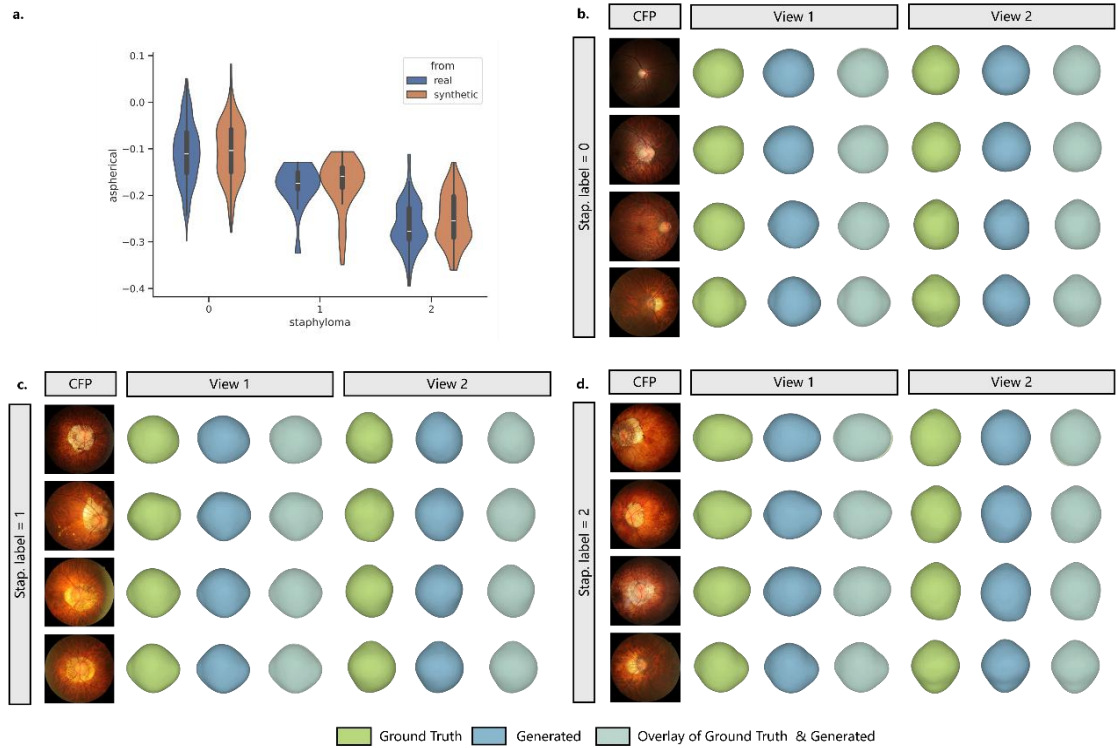

**Fig. 3: Generated eyeballs preserve shape distribution across different types of posterior staphyloma.**

a. Consistency of the aspheric descriptor (Q value) distribution between real and generated 3D eye globe (EG) models across staphyloma subtypes. Q values quantify the asphericity of the globe shape, with preserved distributions indicating accurate geometric replication by the generative model.

b-d. Examples of real and generated 3D EG across different staphyloma types [0 = no staphyloma, 1 = staphyloma affecting the macula, 2 = staphyloma sparing the macula]. View 1 = horizontal, View 2 = vertical, CFP = color fundus photograph.

Finally, we performed external validation on the PALM dataset.[19] The PALM dataset comprises 1200 fundus images, each associated with binary labels for pathological myopia. Specifically, we input only the fundus images into the Fundus2Globe, with all other conditions set to None. For the generated EGs, we assessed the distribution of Q value between the cases with and without pathological myopia. As shown in Fig. 4a, we presented models corresponding to five different training data splits, where the Q-values in the group with pathological myopia are significantly lower than those in the group without pathological myopia ($P < 0.001$). Further breaking down by the official data splits, this trend remains unchanged (Fig. 4b). We showcased the generated EGs from the PALM dataset in Supplementary Fig. 5,6 for qualitative assessment by visual inspection.

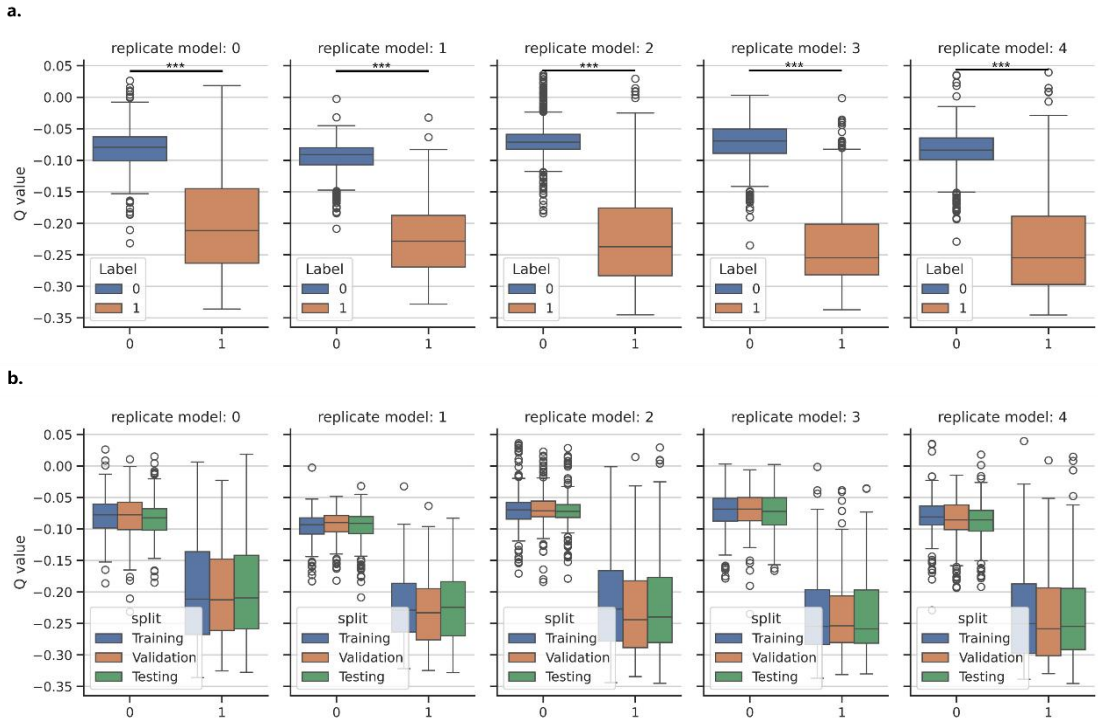

**Fig. 4: External validation on PALM dataset using the shape descriptor Q.**

a. Box plots comparing Q-value distributions between eyes with (label 1) and without (label 0) pathological myopia. Cross-validation across five data splits demonstrates consistent Q values in pathological myopia cases ($P < 0.001$, all splits).

b. Stratification by the PALM dataset's train/validation/test splits confirms the robustness of this trend across independent cohorts, underscoring the generalizability of Fundus2Globe.

**Fundus2Globe generates plausible counterfactuals as personalized digital twins**

Fundus2Globe's ability to simulate hypothetical scenarios was probed via counterfactual generation. In this process, a random CFP was used as a confounder, while the ground-truth metadata (axial length and spherical equivalent) served as factual conditioning. and a set of random yet plausible metadata was used as counterfactual conditioning. Due to the inherent correlation between AL and SE, it is difficult to sample these attributes independently. Therefore, we used an implicit approach to sample both attributes simultaneously. We performed clustering in the joint space of AL and SE and then applied over-sampling using SMOTE[30] to generate samples that are meaningful for observation but not present in the original dataset (Fig. 5a).

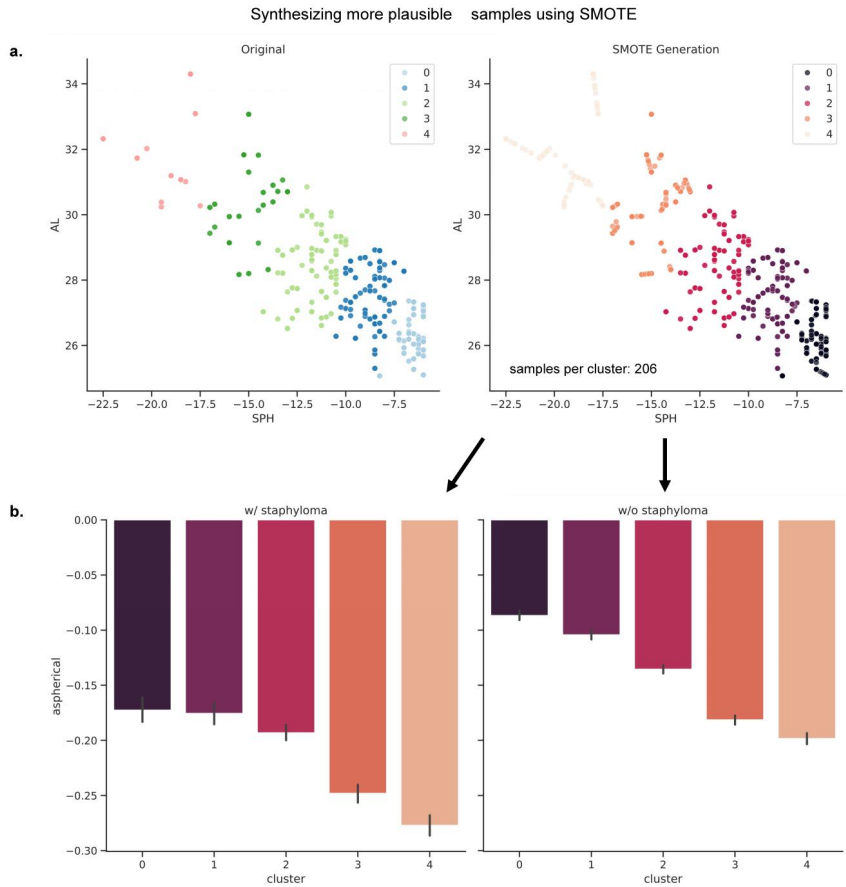

**Fig. 5: Counterfactual generation framework and quantitative shape analysis.**

a. Synthetic sampling of axial length (AL) and spherical equivalent (SE). Clustering was applied to the original AL-SE distribution (left), followed by SMOTE oversampling to generate novel, clinically plausible combinations (right).

b. Counterfactual-generated 3D eye globes exhibit increased asphericity (lower Q-values) with smaller SE and larger AL. Staphyloma-positive cases (left) demonstrate greater shape sensitivity to parameter changes than staphyloma-negative counterparts (right), likely reflecting broader morphological variability in eyes without biomechanical constraints from staphyloma.

As shown in Fig. 5b, the counterfactual conditioning sample points reveal that a smaller SE and a longer AL result in greater asphericity (smaller Q value) in the generated 3D EGs. Additionally, staphyloma-positive cases demonstrated more pronounced shape changes compared to staphyloma-free groups, likely reflecting broader shape variation in the latter due to a larger sampling space. Visualization examples (Fig. 6, Supplementary Figs. 7-8) revealed parameter-driven EG deformations such as barrel-shaped or ellipsoidal geometries under extreme AL/SE values. In Supplementary Figs. 7-8, where atrophic lesions are more localized, the generated EG was more likely to be affected by AL and SE, affecting peripheral shapes. Notably, counterfactual-generated EGs occasionally showed minimal variation in cases with extensive lesions (e.g., large chorioretinal atrophy). For instance, EGs in Supplementary Fig. 9 (rows 3-4) with severe macular atrophy remained saddle-shaped (low AL, high SE) or ellipsoidal/barrel-shaped (high AL, low SE), suggesting lesion-driven constraints on the model's generative sampling space. These findings highlight Fundus2Globe's potential as a digital twin framework while underscoring opportunities to enhance its robustness in complex pathological contexts through expanded training data and scenario diversity.

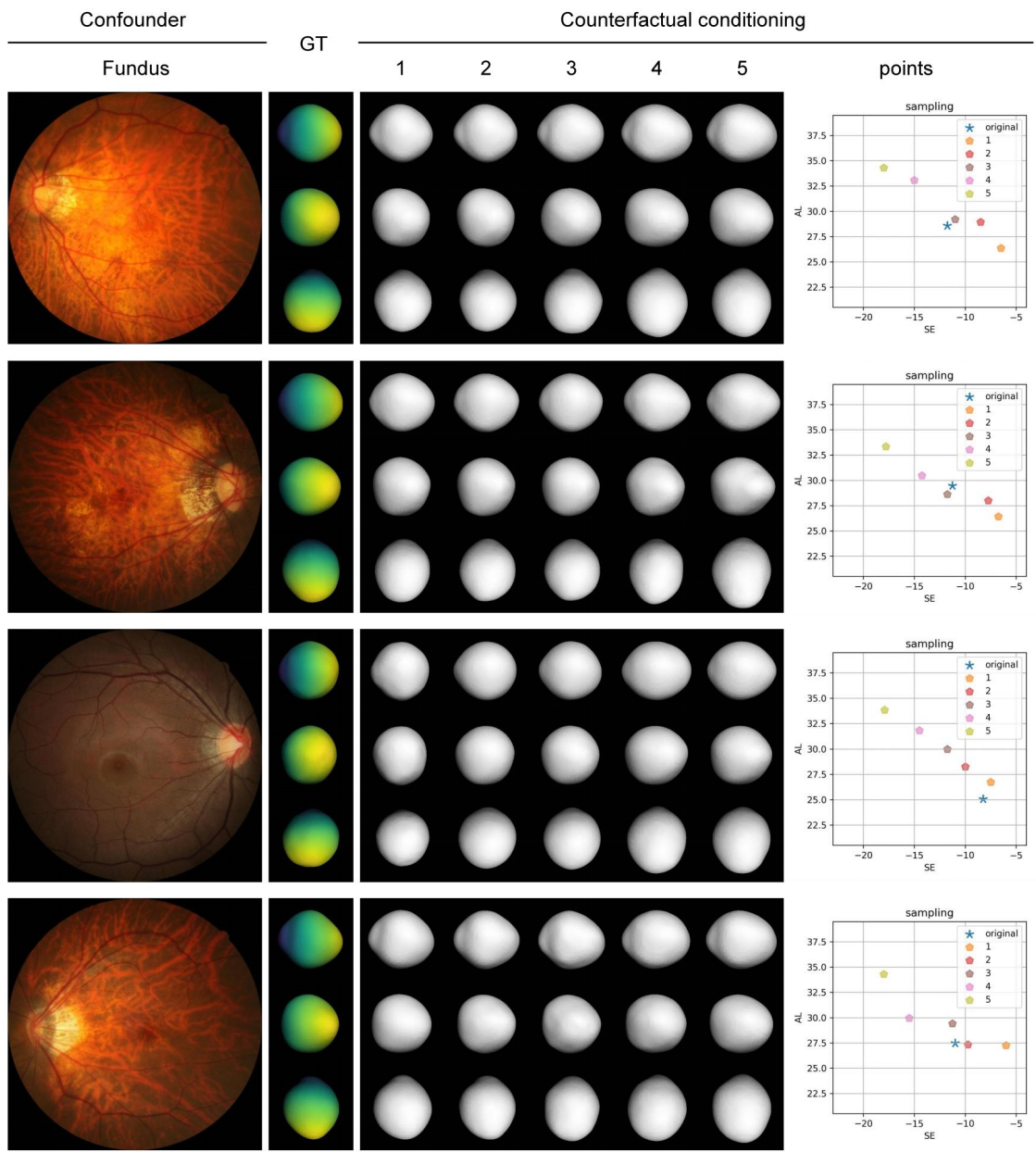

**Fig. 6: Visual inspection of counterfactual generation.**

Examples of shape changes in 3D eye globes during counterfactual generation. In this process, a random color fundus photograph was used as a confounder. The actual metadata (axial length [AL] and spherical equivalent [SE]) served as factual conditioning (labeled as 'original'), while a set of random yet plausible metadata was used for counterfactual conditioning (labeled as '1, 2, 3, 4, 5'). More examples are presented in **Supplementary Fig. 7-9.**

**Discussion**

We present Fundus2Globe, a novel framework for generating 3D EGs that preserve pathological lesions and abnormal ocular shapes using widely accessible CFPs and metadata (AL] and SE). This approach addresses a critical gap in resource-limited settings where high-resolution MRI—the gold standard for ocular shape analysis—is unavailable for routine screening. By synthesizing anatomically plausible EGs, Fundus2Globe provides actionable insights into myopia progression while circumventing costly imaging modalities.

Our findings align with previous studies emphasizing the significance of ocular shape as a biomarker of myopia and its complications.[5-10,31,32] As myopia severity increases, the eyeball shape transitions from an oblate form to spherical and prolate.[9,33] Moreover, pathological myopes exhibit distinct anisotropic and non-uniform expansion patterns.[6] Yu et al. observed that myopic traction maculopathy eyes undergo non-uniform expansion.[8] Guo et al. found barrel-shaped and temporally distorted eye shapes distinguish myopic maculopathy from simple high myopia.[7] Fundus2Globe accurately replicates shape variations observed in pathological myopia, including more pronounced prolate shapes and lower Q values in eyes with posterior staphyloma. This consistency confirms its ability to retain clinically relevant features. By generating patient-specific digital replication of 3D EG, our model may provide early prediction of vision-threatening complications and inform personalized interventions—a capability unattainable with static imaging alone.

The counterfactual generation capability of Fundus2Globe introduces a paradigm shift in mechanistic exploration and clinical hypothesis testing. By simulating hypothetical AL/SE combinations (Fig. 5-6), clinicians can investigate causal relationships between biometric parameters and ocular remodeling. For example, staphyloma-positive EGs showed greater shape sensitivity to AL/SE changes than staphyloma-negative cases, suggesting lesion-driven biomechanical constraints. This aligns with clinical observations where extensive chorioretinal atrophy limits ocular deformability (Supplementary Fig. 9). From a future perspective, counterfactuals may serve as a virtual lab to identify shape-based biomarkers for disease staging, optimize surgical planning (e.g., staphyloma resection), and validate theoretical models of myopic progression.

While MRI remains the gold standard for 3D ocular shape imaging, its cost and accessibility limitations hinder widespread adoption (Table 2).[32] Fundus2Globe bridges this gap by leveraging CFPs—a ubiquitous, low-cost modality—to democratize 3D shape analysis. However, three key opportunities merit further exploration. First, integrating OCT's high-resolution local data with MRI's global shape context[34,35] could enhance anatomical fidelity. Second, optimizing generative models through topology-aware architectures and scalable representations could improve pathological detail and shape variability. Third, expanding counterfactual scenarios—such as simulating intraocular pressure changes or other disease outcomes—would strengthen clinical utility and support personalized therapeutic strategies, which warrant future investigation.

Table 2. The advantages and disadvantages of MRI and Fundus2Globe.

| Method | MRI (Gold Standard) | Fundus2Globe (Proposed) |
| --- | --- | --- |
| **Examination time** | 20–60 minutes | <3 minutes (instant generation from CFP + metadata) |
| **Cost** | High (800–2,000 CNY) | Low (100-300 CNY) |
| **Accuracy** | Gold standard for 3D eye shape | Realistic approximations |
| **Resolution** | 0.1–1 mm³ | Flexible, dependent on training data |
| **Accessibility** | Limited (requires specialized MRI scanners, trained technicians) | High (deployable in clinics with CFP capabilities) |
| **Patient comfort** | Moderate (claustrophobia risk, requires immobilization) | No risk |
| **Dynamic adjustments** | Static (single time-point structure assessment) | Dynamic (generates "what-if" scenarios by modifying clinical parameters) |
| **Clinical usage** | Diagnosis, research, surgical planning | Disease progression simulation, counterfactual analysis, personalized therapy planning |

## Methods

### Data acquisition

The Fundus2Globe model was trained using data from the Zhongshan Ophthalmic Center Brien Holden Vision Institute (ZOC-BHVI) Guangzhou High Myopia Cohort Study, referred to as the internal dataset. This dataset was randomly divided at the participant level into a developmental set (80%) and an internal test set (20%). The developmental set was further partitioned into training and validation subsets, ensuring that all examinations from a single participant remained within the same subset to avoid data leakage. At baseline, the cohort enrolled individuals aged 7–70 years with bilateral high myopia (spherical power ≤ −6.00 D) and no secondary causes of myopia, prior ocular surgeries, or systemic comorbidities. For the MRI substudy, a 10% subsample was selected using stratified random sampling across two age strata (<40 and ≥40 years), excluding individuals with MRI contraindications (e.g., cardiac pacemakers, metallic implants). Of 100 invited participants, 75 were under 40 years old and 25 were 40 years or older.

**Ophthalmic Examinations** All participants underwent standardized ophthalmic evaluations. Best Corrected Visual Acuity (BCVA) was measured using retroilluminated logMAR charts with tumbling-E optotypes (Precision Vision, La Salle, IL). Axial length was quantified using optical low-coherence reflectometry (Lenstar LS900; Haag-Streit AG, Switzerland) for values ≤32 mm, and partial coherence interferometry (IOL Master; Carl Zeiss Meditec, Germany) for longer axial lengths. Cycloplegic refraction was performed with an autorefractor (KR8800; Topcon, Japan) after confirming complete cycloplegia (pupil dilation ≥6 mm, absent light reflex).

**3D MRI Acquisition** Turbo spin-echo T2-weighted MRI scans were acquired at Guangzhou Brain Hospital using a 3.0T Philips Achieva scanner with an 8-channel phased-array head coil. Participants were instructed to keep their eyes closed and minimize movement during the 10-minute scan. Imaging parameters included a field of view of 256 × 256 × 188 mm, isotropic resolution of 1 × 1 × 1 mm, and sequences optimized to enhance fluid-filled ocular structures while suppressing fat signals.

**Reconstruction of 3D EGs**

A two-stage pipeline was developed to reconstruct 3D EGs from MRI data, as detailed in Supplementary Fig. 5. In the segmentation phase, we first generate all mask proposals for each 2D MRI slice. Next, we identified the 3D bounding box of the eye regions in the coronal and transverse planes by leveraging eyeball symmetry and circularity of the mask regions. Masks corresponding to the eyeballs were selected based on the highest Intersection over Union (IoU) value. Each mask was refined using a top-hat filter to remove small objects, followed by connectivity analysis to eliminate over-segmented regions. To mitigate over-segmentation caused by small reflected eye regions in certain slices, we excluded masks with IoU < 0.4 relative to the largest mask in each orientation. The final set of 2D masks was then assembled into a 3D voxel mask.

In the meshing phase, we first generated a triangular mesh using the marching cubes algorithm (via PyMCubes[36]). We then applied Poisson surface reconstruction and midpoint subdivision (PyMeshLab37) to refine the 3D structure. The parameters used were: Poisson surface reconstruction (depth=10, samples per node=5.0, point weight=0.0) and midpoint subdivision (2 iterations).

In the meshing phase, a triangular mesh was generated using the marching cubes algorithm (PyMCubes[36]), followed by Poisson surface reconstruction (depth=10, samples per node=5.0, point weight=0.0) and midpoint subdivision (2 iterations) in PyMeshLab[37] to smooth and refine the 3D structure.

**3D Morphable Model for ocular shape representation**

Operating the diffusion model in the proposed Fundus2Globe directly in mesh space would require prohibitively expensive data resources. To facilitate diffusion model training with limited data while preserving quality, we incorporated strong shape priors. Specifically, we designed a morphable model to represent 3D ocular shapes, enabling encoding and decoding of 3D EGs by transforming between 3D space and an ocular latent space, which involves 3D reconstruction and registration, and morphable model construction.

Before constructing the model, we first registered the 3D ocular shapes using the eyeball model from the FLAME framework[38] as the reference mesh through iterative alignment steps to construct 3D EG meshes: 1. orientation correction: We fitted a spherical surface to the front half of the eyeball to identify the corneal area, then aligned corneal centers to determine the rotation vector; 2. Scaling Adjustment: The transverse length of the eyeball was scaled to match anatomical proportions. 3. initial rigid alignment: The Iterative Closest Point (ICP) algorithm in PyMeshLab refined the fit between the source and reference meshes. 4. fine-tuning via non-rigid ICP: we uniformly sampled 50 landmarks along the corneal edge and adjacent regions, then applied a non-rigid ICP algorithm[39,40] with constraints on the posterior half of the eyeball to prevent excessive deformation. 5. vertex standardization: we oversampled vertices of the reference mesh and used nearest-neighbor search to ensure all meshes contained the same number of vertices.

After registration, each eyeball's geometry was represented as a shape vector: $S = (X_1, Y_1, Z_1, X_2, ..., Y_n, Z_n)^T \in \Re^{3n}$, where $n$=1448 is the number of vertices. A morphable model was then built using $m$=114 exemplar shapes, each defined by its shape vector $S_i$. Any new shape $S_{model}$ was represented as a linear combination of example shapes:

$$S_{model} = \sum_{i=1}^{m} \alpha_i S_i$$

To extract statistical shape properties, Principal Component Analysis (PCA) was performed, reducing the shape space to dimension $d$=$m$-1.[41] The final shape representation was expressed in terms of eigenvectors $s_i$:

$$S_{model} = \bar{S} + \sum_{i=1}^{d} \alpha_i s_i$$

where $\bar{S}$ is the mean shape of the dataset, and each ocular shape was parameterized by

$$\boldsymbol{\alpha} = (\alpha_1, \alpha_2, ..., \alpha_d)^T$$

Separate models were built for left and right eyes.

**Foundation model for CFP representation**

Directly modeling CFP—a high-dimensional imaging modality—is computationally expensive and prone to overfitting. To address this, we leverage EyeFound[25], an ophthalmology foundation model pre-trained on 11 multimodal ophthalmic imaging modalities, to project CFP into a compact latent space. EyeFound employs a ViT-Large architecture trained via self-supervised reconstruction, learning a shared representation across modalities. This approach achieves performance comparable to (or exceeding) modality-specific encoders while enabling unified multi-modality task adaptation with a single image encoder. By distilling cross-modal anatomical priors, EyeFound encodes CFP into a 1024-dimensional latent vector that captures clinically meaningful features, dramatically reducing computational complexity for downstream tasks.

**Fundus2Globe Architecture and Training**

Fundus2Globe is a diffusion model trained in the ocular latent space to generate 3D eye meshes. Diffusion models are probabilistic generative models that iteratively denoise Gaussian-distributed samples via a learned reverse Markov chain process, comprising a forward process that gradually corrupts input data to Gaussian noise over T timesteps and a reverse process: a U-Net denoiser trained to reconstruct the original data from noisy inputs.[42,43]

For a given ocular latent encoded by EyeFound, we applied a forward diffusion process over 1000 timesteps, progressively adding Gaussian noise until the latent distribution approximated isotropic Gaussian distribution. The denoising U-Net[44] was trained with a dual objective: (1) predicting the original latent representation by minimizing the mean squared error (MSE) between its output and the noise-free input, and (2) ensuring geometric fidelity by minimizing the Chamfer distance between the decoded 3D mesh and the ground-truth triangle mesh. The U-Net contains four residual downsampling blocks[45] (channel dimensions: 256, 256, 512, 1024), designed to process spatialized latent representations. To adapt the $d$-dimensional latent vector for the U-Net's convolutional layers, we reshaped and tiled it into a $d \times h \times w$ tensor, effectively creating a spatial feature map. During inference, accelerated sampling (10 steps) generated denoised outputs, which were condensed back into a $d$-dimensional vector via spatial average pooling. Finally, the refined latent code was decoded into a 3D mesh using the 3DMM optimized for ophthalmic anatomy.

For the conditional guidance, the fundus embedding extracted from the pre-trained encoder EyeFound was a 1024-dimensional vector. The left or right eye label was encoded using a simple lookup table implemented with torch.nn.Embedding in PyTorch[46], mapping the label to a 1024-dimensional vector. AL and SE were discretized separately using label distribution learning (LDL), converting each into a 512-dimensional vector. These two vectors were concatenated into a single 1024-dimensional vector. In the LDL-based discretization method, each value was transformed into a Gaussian-distributed vector with a mean equal to the given label value and a standard deviation of 1.0. The three resulting 1024-dimensional vectors (fundus embedding, eye label, and discretized biometric parameters) were element-wise summed, and the final 1024-dimensional vector was combined with the time-step condition to guide the generation process.

The encoder and decoder of Fundus2Globe were kept frozen while the diffusion model was trained on an NVIDIA A800 GPU for 6020 steps with a batch size of 16. We used the Adam optimizer with a learning rate of 1e-4, and the probability of condition dropout during training was set to 0.5.

**Evaluation Metrics**

To assess the accuracy and realism of the synthetic EGs, we utilized three evaluation metrics: Chamfer Distance, Hausdorff Distance, and Point-to-Surface Distance.

The Chamfer Distance was used as the primary metric to quantify topological differences between the synthetic EGs and their realistic counterparts. It is defined as:

$$d_{CD}(S_1, S_2) = \frac{1}{S_1} \sum_{x \in S_1} \min_{y \in S_2} \|x - y\|_2^2 + \frac{1}{S_2} \sum_{y \in S_2} \min_{x \in S_1} \|y - x\|_2^2$$

where $S_1$ and $S_2$ represent two sets of 3D point clouds. The first term computes the sum of minimum squared Euclidean distances from each point in $S_1$ to $S_2$, while the second term does the same in the opposite direction. $\|\sim\|_2$ denotes the usual Euclidean norm

The Hausdorff Distance measures the worst-case discrepancy between two meshes and is defined as:

$$d_H(S_1, S_2) = \max \left\{ \sup_{x \in S_1} \inf_{y \in S_2} d(x, y), \sup_{y \in S_2} \inf_{x \in S_1} d(x, y) \right\}$$

where $d(x, y)$ denotes the Euclidean distance between two points. This metric captures the maximum deviation between the two surfaces, ensuring that both large-scale structural differences and localized inconsistencies are accounted for.

The Point-to-surface distance was used to analyze the spatial distribution of generation quality, measuring how well the synthetic surface aligns with the real one:

$$d(p, S') = \min_{p' \in S'} \|p - p'\|_2$$

where $p$ is a point on the surface $S$, $p'$ is a point on the surface $S'$. This metric evaluates the fidelity of generated meshes by ensuring that each point on one surface has a close corresponding point on the other.

**Ethics Statement**

This study was conducted in accordance with the Declaration of Helsinki and received approval from the Hong Kong Polytechnic University's institutional review board (HSEARS20240202004). The IRBs waived informed consent due to the retrospective analysis of anonymized ophthalmic images.

**Competing interests**

A patent has been filed for this innovation: "Method and System for 3D generation from 2D imaging. 202410360491.4"

**Data availability**

We do not have permission to redistribute the internal dataset. More details of the PALM dataset can be found at https://doi.org/10.6084/m9.figshare.21299148.v1.

**Code availability**

The code is available at https://github.com/xxx, which is based on PyTorch.

**Author Contributions**

D.S. was involved in the conceptualization and design of the study.

D.S. and O.X. played a role in acquiring and organizing the data.

B.L. was responsible for the technical execution of the work.

M.H. offered clinical insights for the research.

J.Y. provided guidance on the analysis framework.

All authors participated in drafting and revising the manuscript.

**Funding**

The study was supported by the Global STEM Professorship Scheme (P0046113) and Henry G. Leong Endowed Professorship in Elderly Vision Health. The sponsors or funding organizations had no role in the design or conduct of this research.

**Acknowledgment**

We thank the InnoHK HKSAR Government for providing valuable support.